\documentclass[]{spie}  

\usepackage{amsmath,amsfonts,amssymb}
\usepackage{graphicx,tabularx,booktabs,parskip}
\usepackage[colorlinks=true, allcolors=blue]{hyperref}

\title{Investigation of unsupervised and supervised hyperspectral anomaly detection}

\author[a]{Mazharul Hossain}
\author[b]{Aaron Robinson}
\author[a]{Lan Wang}
\author[b]{Chrysanthe Preza}
\affil[a]{Computer Science Department}
\affil[b]{Electrical and Computer Engineering Department}
\affil[ ]{The University of Memphis, Memphis, TN, USA}

\authorinfo{Further author information: (Send correspondence to Mazharul Hossain); E-mail: mhssain9@memphis.edu}

\begin{document} 
\maketitle

\begin{abstract}
Hyperspectral sensing is a valuable tool for detecting anomalies and distinguishing between materials in a scene. Hyperspectral anomaly detection (HS-AD) helps characterize the captured scenes and separates them into anomaly and background classes. It is vital in agriculture, environment, and military applications such as RSTA (reconnaissance, surveillance, and target acquisition) missions. We previously designed an equal voting ensemble of hyperspectral unmixing and three unsupervised HS-AD algorithms. We later utilized a supervised classifier to determine the weights of a voting ensemble, creating a hybrid of heterogeneous unsupervised HS-AD algorithms with a supervised classifier in a model stacking, which improved detection accuracy. However, supervised classification methods usually fail to detect novel or unknown patterns that substantially deviate from those seen previously. In this work, we evaluate our technique and other supervised and unsupervised methods using general hyperspectral data to provide new insights.
\end{abstract}



\section{Introduction}
\label{sec:intro}  

Hyperspectral remote sensing is a technique that combines spectroscopy and imaging to collect spatial and spectral information and process information from across the electromagnetic spectrum. 
Hyperspectral imaging (HSI) aims to obtain the spectral information from a scene, and advancement of HSI technologies~\cite{khan2018modern} combined with machine learning (ML) recognition techniques~\cite{paoletti2019deep} promises improved scene characterization in numerous applications, such as the detection of anomalies in a large background, find objects, or identify materials.
{\bf {Hyperspectral anomaly detection (HS-AD)} }\cite{su2021hyperspectral} helps characterize the captured scenes and separates them into anomaly and background classes. Supervised anomaly detection solves this problem using data labeled with ground truth, whereas unsupervised anomaly detection does not.

Universality or generalization suggests that if a method achieves promising results on known datasets, it will perform similarly well for new or unseen data and provide consistent predictions. Usually, unsupervised HS-AD methods promise generalization as they should work well enough on unknown data. 
We previously introduced a new approach~\cite{younis2023hyperspectral, hossain2024greedy} to the HS-AD problem using a supervised classifier as a meta-model in a stacking~\cite{hossain2024greedy} ensemble, which improved detection accuracy. However, our previous method does not generalize well when trained on one dataset and tested on other unfamiliar datasets. Unlike supervised ML models, unsupervised HS-AD methods should work well enough on unknown data. Unfortunately, our evaluation found unsupervised HS-AD methods are not consistently generalizing, and their performance varies from dataset to dataset.

This treatment presents our unsupervised stacking-ensemble anomaly detection approach to solve the generalization problem. We have utilized the Greedy search~\cite{hossain2024greedy} approach to find suitable HS-AD methods for the base model and unsupervised Gaussian mixture model as the meta-model. We believe this approach holds significant potential for improving generalized anomaly detection accuracy. 
We have considered twenty-two scenes from four public and one private benchmarking datasets (thirteen scenes from the ABU (airport, beach, and urban) dataset, five from the Arizona dataset, two from the San Diego airport dataset, and one from each of the HYDICE urban and Salinas datasets). We used 2-fold cross-validation five times to show our methods' statistical significance, where it achieved an average ROC-AUC score of 0.918 with a standard deviation of 0.131, which is at least seven \% better than other individual methods. This article also investigates other supervised and unsupervised methods using these datasets. We compare the ROC-AUC scores to provide new insights and recommendations on the performance. Continuous improvement and innovation in this research area are crucial in further enhancing the general accuracy of hyperspectral anomaly detection and advancing the hyperspectral sensing field.


\section{Background}

Hyperspectral imagers are expensive, and only a few standard HS datasets are available~\cite{hu2022hyperspectral}. As a result, most researchers are exploring unsupervised HS-AD algorithms~\cite{xu2022hyperspectral} that do not require annotated data, such as those based on statistical methods~\cite{su2021hyperspectral}.


\subsection{Individual Hyperspectral Anomaly Detection~Algorithms}

The researchers and reviewers categorize hyperspectral anomaly detection (HS-AD) methods based on the design criteria to separate the background and anomalies and also based on the primary function model~\cite{raza2022hyperspectral, su2021hyperspectral, xu2022hyperspectral}. The Reed--Xiaoli (RX)~\cite{reed1990adaptive, chang2002anomaly}, MD--RX, and Windowed RX (WIN-RX) are some examples of Statistics-based HS-AD techniques. 
Subspace-RX (SSRX)~\cite{schaum2002joint}, Complementary Subspace Detector (CSD)~\cite{schaum2007hyperspectral}, Local Summation Anomaly Detection (LSAD)~\cite{du2016spectral}, Local Summation Unsupervised Nearest Regularized Subspace with an Outlier Removal Anomaly Detector (LSUNRSORAD)~\cite{tan2019anomaly} are some examples of subspace-based techniques. 
Kernel--RX Algorithm (KRX)~\cite{kwon2005kernel, hidalgo2020efficient}, Gaussian Mixture RX (GM-RX)~\cite{acito2005gaussian} are some examples of kernel-based techniques. 
Cluster-based anomaly detector (CBAD)~\cite{carlotto2005cluster}, fuzzy c-means clustering-based anomaly detector (FCBAD)~\cite{hytla2009anomaly} are some examples of clustering distance-based techniques. 
Attribute and Edge-Preserving Filters (AED)~\cite{kang2017hyperspectral} is an example of spatial-spectral filtering-based techniques. 
Isolation forest (iForest)~\cite{liu2008isolation} is an example of Machine learning-based techniques. 
Kernel iForest (KIFD)~\cite{li2019hyperspectral} combines both kernel-based and ML-based techniques.
We found that each method's performance depends upon specific background characteristic constraints and is effective only in some scenarios. Thus, we investigated to find a combination of methods to overcome this limitation.


\subsection{Ensemble Hyperspectral Anomaly Detection~Algorithms}

\begin{figure}[!htbp]
    \centering 
    \includegraphics[width=0.8\linewidth]{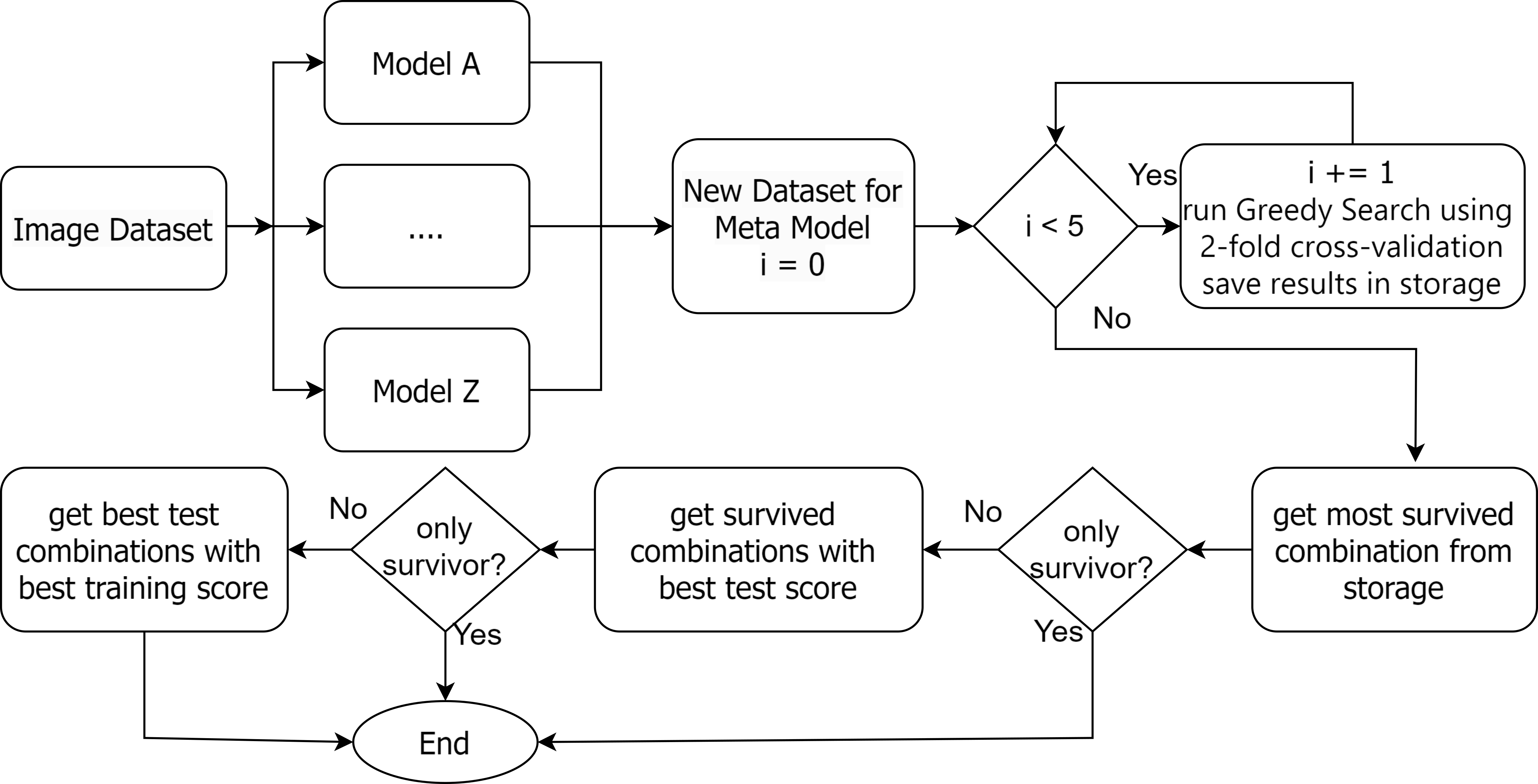}
    \caption{\label{fig:GE_algo_diagram}Overall flowchart of GE-AD~\cite{hossain2024greedy} to find the best AD methods for ensemble~fusion.}
\end{figure}

Yang~et~al.~\cite{yang2022ensemble} proposed an Ensemble and Random RX with Multiple Features (ERRX MF) anomaly detector using three features, Gabor, Extended Morphological Profile (EMP), and Extended Multiattribute Profile (EMAP), along with the original hyperspectral image.
Wang~et~al.~\cite{wang2022subfeature} proposed a sub-feature ensemble called SED. They randomly sub-sampled channels and used six heterogeneous base models in the ensemble. Finally,  a prior-based tensor approximation algorithm (PTA)~\cite{li2020prior} used these sub-features for anomaly detection.
Fatemifar~et~al.~\cite{fatemifar2020stacking} proposed a stacking ensemble for face spoofing anomaly detection that consisted of 63 base classifiers and a Gaussian Mixture Model (GMM) as the meta-classifier for the second stage. 
Younis~et~al.~\cite{younis2023hyperspectral} proposed an equal-weight voting method called Hyperspectral Unmixing-Based Voting Ensemble Anomaly Detector (HUE-AD) that combines four detectors (Abundance, AED, KIFD, and LSUNRSORAD) to identify anomalies. 
Hossain~et~al.~\cite{hossain2024greedy} proposed a stacking ensemble called Greedy Ensemble Hyperspectral Anomaly Detector (GE-AD) that uses a greedy search algorithm to find four suitable base detectors and utilizes a supervised machine learning model to combine those four base detectors to identify anomalies. Figure~\ref{fig:GE_algo_diagram} shows the overall flowchart of the GE-AD algorithm. The greedy search algorithm uses 2-fold cross-validation to evaluate performance metrics and greedily searches for base methods to maximize the performance score, which reduces the search space compared to a grid search approach. This method produces a much higher F1-macro score and similar ROC-AUC score than the input methods.

\vspace{-6pt}


\section{Methodology}

We optimized the HUE-AD and GE-AD in this treatment to improve their generalizability. We also investigated some unsupervised ML models as meta-models. Unlike the F1 score, we found that improving the ROC-AUC score in one dataset translates to a better ROC-AUC score in another. Thus, we tried to maximize the ROC-AUC score using the greedy search. In this section, we discussed our methodology in detail.


\subsection{Hyperspectral Unmixing-Based Voting Ensemble Anomaly Detector (HUE-AD)}

Our empirical tests found that HUE-AD~\cite{younis2023hyperspectral}'s three base anomaly detectors can operate adequately in a trio even when the particular spectra of interest or their abundances are unknown. Thus, under our new modification, we skip the abundance and determine a pixel as anomalous if at least two base methods vote a pixel as anomalous. Although not as good as the original HUE-AD, the performance was acceptable and still has a higher F1-macro score than the input methods. As it has a lower ROC-AUC score, we modified it further and used an average ensemble using the input results without binarizing to find the anomaly score. We are calling it the baseline in our investigation.


\subsection{Greedy Ensemble Hyperspectral Anomaly Detector (GE-AD)}

GE-AD~\cite{hossain2024greedy} is a stacking ensemble that uses a supervised machine learning model as a meta-model that combines four detectors to identify anomalies. GE-AD utilizes a greedy search algorithm to find those four suitable detectors to maximize the scoring function. However, the trained model does not generalize and suffers when applied to an unfamiliar new scene. Thus, along with four base methods, we used thirty normalized random channels as input to the supervised meta-classifier. We expect the classifier to learn about the underlying dataset and generalize better. Our results show better ROC-AUC scores than before. We are calling this modified version the mGE-AD in our investigation.


\subsection{Proposed Unsupervised Ensemble Approach (UGE-AD)}

We have utilized the same greedy search~\cite{hossain2024greedy} approach to find suitable HS-AD methods as the base model. We considered the Reed--Xiaoli (RX)~\cite{reed1990adaptive, chang2002anomaly} detector, Attribute and Edge-Preserving Filters (AED)~\cite{kang2017hyperspectral}, Cluster-based anomaly detector (CBAD)~\cite{carlotto2005cluster}, Complementary Subspace Detector (CSD)~\cite{schaum2007hyperspectral}, Fuzzy clustering-based anomaly detector (FCBAD)~\cite{hytla2009anomaly}, Gaussian mixture RX (GM-RX)~\cite{acito2005gaussian} anomaly detector, Kernel iForest (KIFD)~\cite{li2019hyperspectral} anomaly detector, Local Summation Unsupervised Nearest Regularized Subspace with an Outlier Removal Anomaly Detector (LSUNRSORAD)~\cite{tan2019anomaly}, Median RX (MD-RX) anomaly detector, Subspace-RX (SSRX)~\cite{schaum2002joint}, Windowed RX (WIN-RX). 

The RX~\cite{reed1990adaptive, chang2002anomaly} detector has a simple statistical principle, low computational complexity, and relatively good performance. It characterizes the background using the HSI's mean and covariance and calculates the Mahalanobis distance~\cite{chandra1936generalised} between the background and pixel under test. As the mean is very susceptible to noise, MD-RX uses the median instead of the mean as statistical information to improve RX. Meanwhile, WIN-RX uses a sliding window to calculate local mean and covariance and determine anomalies.
GM-RX~\cite{acito2005gaussian} uses the Gaussian Mixture Model (GMM) to characterize the complex scene to improve the detection of the RX algorithm. 
Schaum~et~al.~\cite{schaum2002joint} in SSRX deleted several high-variance Principal Components (PC) dimensions as background clutters found in principal component analysis (PCA) to improve the RX performance. They later designed CSD~\cite{schaum2007hyperspectral} where the highest variance PCs define the background subspace and the others (the complementary subspace) as the target subspace.
Carlotto, Mark J.~\cite{carlotto2005cluster} proposed that the image background can be divided into clusters in the CBAD algorithm. FCBAD~\cite{hytla2009anomaly} improved the CBAD using fuzzy c-means clustering instead of k-means clustering.
Kang~et~al.~\cite{kang2017hyperspectral} assumed that anomalies tend to be smaller and possess unique reflectance signatures and that the pixels belonging to the same class would have a high correlation in the spatial domain, and based on this principle, they proposed AED.
Tan~et~al.~\cite{tan2019anomaly} proposed LSUNRSORAD to decrease the computational complexity of LSAD~\cite{du2016spectral}. 
Li~et~al.~\cite{li2019hyperspectral} utilized the kernel space of PCA and the isolation forest (iForest)~\cite{liu2008isolation} to isolate anomalies in their proposed KIFD~\cite{li2019hyperspectral} algorithm.

We considered a similar idea to Wang~et~al.~\cite{wang2022subfeature} and Fatemifar~et~al.~\cite{fatemifar2020stacking} and investigated multiple unsupervised methods as meta-model. We used the Gaussian mixture model for the meta-model. We also investigated Forest; however, it did not perform well.

We have considered the proposed methodology of Yang~et~al.~\cite{yang2022ensemble}. 
Instead of directly passing through the HSI data, we utilized PCA to find ten PCs as a passthrough to the meta-model to improve the ROC-AUC score.


\subsection{Datasets}

We assessed the AD algorithms' effectiveness using twenty-two scenes from six public and one private benchmarking datasets. These scenes are from the Airport--Beach--Urban (ABU) dataset~\cite{kang2017hyperspectral, kang_abu_airport}, the~San Diego dataset~\cite{zhao2014hyperspectral, zhu2018hyperspectral}, the~Salinas dataset~\cite{salinas_data}, the~Hydice Urban dataset~\cite{hydice_data, kalman1997classification}, and~the Arizona dataset~\cite{watson2023evaluation}.

The ABU airport and beach datasets have four scenes, whereas the ABU urban and the Arizona datasets have five. The San Diego airport dataset has two scenes. Both the HYDICE urban and the Salinas datasets have one scene each.
Salinas dataset contains a farming field, whereas HYDICE urban has some urban development within vast vegetation. The ABU urban contains urban scenes. Arizona dataset contains a sandy desert scene with sparse vegetation and no urban development. The ABU beach contains scenes with water, which is challenging because of it. The ABU airport and the San Diego airport contain similar scenes from airports.


\section{Results}

Our research aims to assist the decision-making processes of the human in the loop. Therefore, our performance goal was to identify all anomalies accurately and minimize the misclassification of positive instances as negative because humans can overlook the misclassification of a negative case as positive. To evaluate the HS-AD algorithms, we utilized the traditional classification evaluation metric known as the area under the receiver operating characteristic (ROC) curve (ROC-AUC) score~\cite{hanley1982meaning, 9909988, rasharmaravindrasharma_2022}, which is suitable for assessing this scenario.

For our first evaluation, we used only one dataset in the greedy search to identify suitable base methods for UGE-AD, aiming to maximize the ROC-AUC score. The results are shown in Table~\ref{tab:001}. Figure~\ref{fig:uge_ad_1_vs_others} helps us understand the results better than the table format. The results show that the greedy search may fail to find the optimal base methods for smaller datasets, and the results generalize more effectively for larger datasets. Table~\ref{tab:001} shows that the performances are correlated to input base methods instead of the scene similarity. As we can see, GM-RX helps get a suitable performance score for the San Diego airport dataset. Another issue was that greedy search could find a better score for the Arizona and Salinas datasets. However, only GM-RX was sufficient for the Arizona dataset, as shown in Table~\ref{tab:001}.

\renewcommand\tabularxcolumn[1]{m{#1}} 
\newcolumntype{Y}{>{\centering\arraybackslash}X}

\begin{table}[!htbp]
\centering
\caption{\label{tab:001}Average ROC-AUC scores of our newly proposed Unsupervised ensemble UGE-AD over various datasets. The first column shows the dataset used in the greedy search to find suitable base methods, and the second column shows those base methods. Where the dataset names in rows and columns match, those cells show the training score after the greedy search.}

    \begin{tabularx}{\textwidth}{@{} Y >{\centering\arraybackslash}m{0.20\linewidth} YYYYYYY @{}}\toprule
    
     & & HYDICE &  Salinas &  San Diego &  ABU airport &  ABU beach &  ABU urban &  Arizona  \\ \midrule
    HYDICE &  	 WIN-RX                        & \textbf{0.997} &  0.951 &  0.962 &  0.943 &  0.963 &  0.964 &  0.802 \\ \midrule
    Salinas & LSUNRSORAD, RX                   & 0.995 &  0.999 &  0.965 &  0.949 &  0.971 &  0.971 &  0.780 \\ \midrule
    San Diego &  	 GM-RX                     & \textbf{0.997} &  0.997 &  0.964 &  0.934 &  0.961 &  0.962 &  \textbf{0.847} \\ \midrule
    ABU airport & AED, GM-RX, KIFD, LSUNRSORAD & 0.995 &  \textbf{1.000} &  0.971 &  \textbf{0.962} &  \textbf{0.985} &  0.966 &  0.712 \\ \midrule
    ABU beach & KIFD, LSUNRSORAD, RX, WIN-RX   & 0.996 &  0.997 &  0.965 &  0.960 &  0.973 &  \textbf{0.976} &  0.766 \\ \midrule
    ABU urban & AED, KIFD, LSUNRSORAD, RX      & 0.996 &  \textbf{1.000} &  0.970 &  \textbf{0.962} &  \textbf{0.985} &  0.971 &  0.706 \\ \midrule
    Arizona & CBAD, FCBAD, GM-RX, KIFD      & 0.994 &  0.999 &  \textbf{0.972} &  0.948 &  0.967 &  0.966 &  0.779 \\ \midrule
                                                                    
    \end{tabularx}

\end{table}

\begin{figure}[htbp]
    \centering
    \includegraphics[width=0.95\linewidth]{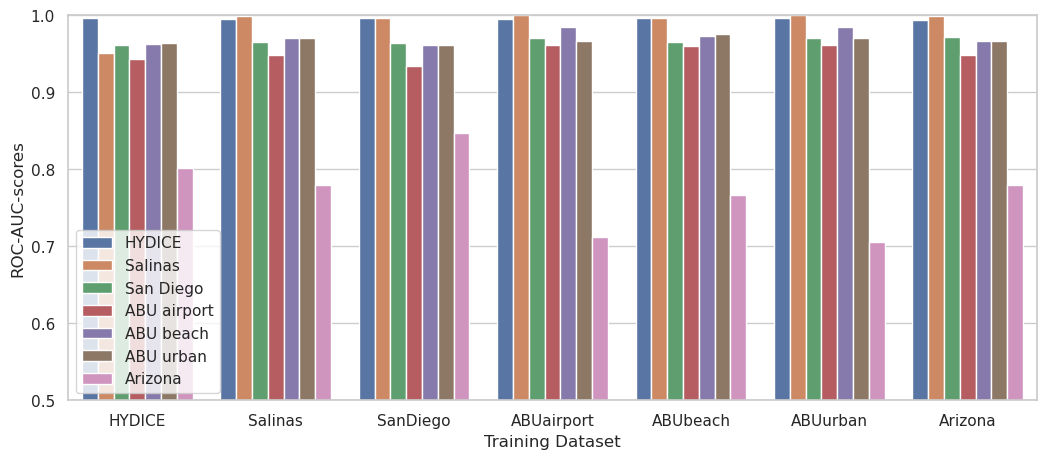}
    \caption{\label{fig:uge_ad_1_vs_others} Average ROC-AUC scores of our UGE-AD over various datasets showing consistent performance against various training datasets. The X-axis shows the dataset used in the greedy search to find suitable base methods. The Y-axis shows the testing ROC AUC scores. Except where the dataset names in rows and columns match, those bars show the training score after the greedy search.}
    
\end{figure}

For our second evaluation, from the twenty-two scenes from six public and one private benchmarking datasets, the GE-AD algorithm randomly selected datasets to ensure we have at least 50\%
image scenes are assigned to the greedy search. Over five runs with 2-fold cross-validation, greedy search tried to maximize the ROC-AUC score and selected AED, FCBAD, GM-RX, and KIFD as base methods for our unsupervised GE-AD. 
In our unsupervised GE-AD, we used these four base methods and ten PCs as inputs to the GMM meta-model. UGE-AD achieved an average of 0.943 ROC-AUC scores with a standard deviation (STD) of 0.088, which is lower than other input methods, as shown in Table~\ref{tab:00} over seven datasets. 
The greedy search selected GM-RX, KIFD, LSUNRSORAD, and MD-RX as base methods for our supervised mGE-AD. We used these four base methods and thirty random channels from the HSI as inputs to the random forest (RF) meta-classifier. mGE-AD achieved an average of 0.882 ROC-AUC score with an STD of 0.170.
Our baseline HUE-AD did not need any training. It is an averaging ensemble of AED, KIFD, and
LSUNRSORAD achieved an average of 0.902 ROC-AUC score with an STD of 0.201.
Table~\ref{tab:01} shows F1-macro scores of these three ensembles. mGE-AD achieved the best average F1-macro score of 0.593 with an STD of 0.070. Although UGE-AD had the lowest STD of 0.048, suggesting a stable solution compared to others, there is still room for improvement.

The results from San Diego airport, HYDICE urban, and Salinas datasets are in Table~\ref{tab:1}. UGE-AD has the highest average in the HYDICE urban dataset. However, LSUNRSORAD and our baseline HUE-AD achieved a perfect score in the Salinas dataset. The baseline HUE-AD scores are better than others for the first scene of the San Diego airport dataset, whereas AED performed best in the second scene.
The individual results from the ABU airport dataset are shown in Table~\ref{tab:2}, where LSUNRSORAD in the first scene, AED performed best in the second and fourth scenes, and baseline HUE-AD in the third scene.
The individual results from the ABU beach dataset are shown in Table~\ref{tab:3}, where AED performed best in the first scene, baseline HUE-AD in the second scene, and UGEE-AD in the fourth scene. However, most of the algorithms got perfect scores in the third scene.
The individual results from the ABU urban dataset are shown in Table~\ref{tab:4}, where AED performed best in the first and third scenes, RX in the second scene, CSD and MD-RX along with RX in the fourth scene, and baseline HUE-AD in the fifth scene.
Table~\ref{tab:5} shows the Arizona dataset results, where GM-RX performed best in the fourth scene and UGE-AD in all four other scenes. 

From our results, it is evident that AED gets better results most of the time. Along with KIFD and LSUNRSORAD, it performs better consistently. Our baseline HUE-AD showed more promising results than UGE-AD. However, UGE-AD showed its performance in the Arizona dataset, where all other HS-AD methods failed. It performed well as the meta-model had access to PCs as input, which helped avoid the pitfall when input base models failed.

\begin{table}[!htbp]
\centering
\caption{\label{tab:00}Average and standard deviation of ROC-AUC scores of various methods computed over the ABU, Arizona, HYDICE urban, Salinas, and San Diego datasets. }

    \begin{tabularx}{\textwidth}{@{} >{\centering\arraybackslash}m{0.18\linewidth} YYYYYYY >{\centering\arraybackslash}m{0.11\linewidth} @{}}\toprule
    
    Methods        & HYDICE urban & Salinas & San Diego & ABU airport & ABU beach & ABU urban & Arizona & Average score \\ \midrule
    CBAD           & 0.970        & 0.390   & 0.611     & 0.753       & 0.681     & 0.466     & 0.566   & $0.634 \pm 0.178$ \\ \midrule
    CSD            & 0.982        & 0.788   & 0.906     & 0.877       & 0.962     & 0.979     & 0.451   & $0.849 \pm 0.175$ \\ \midrule
    FCBAD          & 0.994        & 0.841   & 0.964     & 0.910       & 0.973     & 0.904     & 0.508   & $0.870 \pm 0.156$ \\ \midrule
    GM-RX          & 0.957        & 0.546   & 0.388     & 0.599       & 0.901     & 0.651     & 0.497   & $0.648 \pm 0.194$ \\ \midrule
    MD-RX          & 0.986        & 0.817   & 0.923     & 0.869       & 0.961     & 0.977     & 0.453   & $0.855 \pm 0.174$ \\ \midrule
    RX             & 0.985        & 0.807   & 0.915     & 0.886       & 0.962     & 0.979     & 0.451   & $0.855 \pm 0.175$ \\ \midrule
    SSRX           & 0.747        & 0.848   & 0.802     & 0.758       & 0.925     & 0.848     & 0.589   & $0.788 \pm 0.099$ \\ \midrule
    WIN-RX         & 0.749        & 0.798   & 0.705     & 0.807       & 0.934     & 0.857     & 0.541   & $0.770 \pm 0.116$ \\ \midrule
    AED            & 0.979        & 0.956   & 0.988     & 0.979       & 0.983     & 0.923     & 0.392   & $0.886 \pm 0.203$ \\ \midrule
    KIFD           & 0.830        & 0.995   & 0.991     & 0.964       & 0.948     & 0.962     & 0.652   & $0.906 \pm 0.116$ \\ \midrule
    LSUNRSORAD     & 0.994        & 1.000   & 0.945     & 0.964       & 0.989     & 0.973     & 0.353   & $0.888 \pm 0.219$ \\ \midrule
    baseline HUE-AD (AED, KIFD, LSUNRSORAD) & 0.962        & 1.000   & 0.988     & 0.980       & 0.986     & 0.987     & 0.411   & $0.902 \pm 0.201$ \\ \midrule
    mGE-AD (GM-RX, KIFD, LSUNRSORAD, MD-RX), RF & 0.927        & 0.997   & 0.939     & 0.907       & 0.965     & 0.970     & 0.472   & $0.882 \pm 0.170$ \\ \midrule
    UGE-AD (AED, FCBAD, GM-RX, KIFD), GMM & 0.995        & 0.999   & 0.970     & 0.960       & 0.980     & 0.968     & 0.731   & $0.943 \pm 0.088$ \\ \bottomrule
    
    \end{tabularx}
\end{table}

\begin{table}[!htbp]
\centering
\caption{\label{tab:01}Average and standard deviation of F1-macro scores of our ensemble methods computed over the ABU, Arizona, HYDICE urban, Salinas, and San Diego datasets. }

    \begin{tabularx}{\textwidth}{@{} >{\centering\arraybackslash}m{0.18\linewidth} YYYYYYY >{\centering\arraybackslash}m{0.12\linewidth} @{}}\toprule
    
    Methods        & HYDICE urban & Salinas & San Diego & ABU airport & ABU beach & ABU urban & Arizona & Average score \\ \midrule

    baseline HUE-AD & 0.519        & 0.512   & \textbf{0.619}     & \textbf{0.629}       & 0.564     & \textbf{0.663}     & 0.541   & $0.578 \pm 0.055$ \\ \midrule
    mGE-AD          & \textbf{0.525}        & \textbf{0.729}   & 0.560     & 0.584       & \textbf{0.628}     & 0.624     & 0.503   & $0.593 \pm 0.070$  \\ \midrule
    UGE-AD          & 0.500        & 0.480   & 0.559     & 0.566       & 0.514     & 0.605     & \textbf{0.613}   & $0.548 \pm 0.048$ \\ \bottomrule

    \end{tabularx}
\end{table}


\section{Conclusion}

Our research showed that various approaches do not always give generalized results, as demonstrated by the higher standard deviation (STD). Our new stack ensemble method with an unsupervised GMM meta-model shows generalization and more reliable results with less variation through better average and lower STD ROC AUC score, which is a positive development. We plan to explore it further in future research.
This unsupervised method provides a better ROC AUC score. In contrast, the stack ensemble method with a supervised Random Forest meta-model ensures a better F1-macro score. We plan to design an HS-AD method that enhances ROC AUC and F1 scores.
We plan to investigate other supervised and unsupervised methods as meta-models and evaluate the impact of various thresholding methods and scoring functions on performance.

\appendix

\section{Quantitative Evaluation}
\label{sec:app_qual_eval}

Detailed quantitative results will help us understand the performance of this anomaly detection model. Tables~\ref{tab:1}, \ref{tab:2}, \ref{tab:3}, \ref{tab:4}, and \ref{tab:5} present prediction metrics for individual methods and our ensemble methods for each HS image across all datasets.  

\begin{table}[!htbp]
\centering
\caption{\label{tab:1}Comparison of ROC-AUC scores between various methods using the HYDICE urban, Salinas, and San Diego datasets}

    \begin{tabularx}{\linewidth}{@{} Y YYYY @{}} \toprule
    
    \textbf{Methods}              & \textbf{HYDICE urban} & \textbf{Salinas} & \textbf{San Diego-01} & \textbf{San Diego-02} \\ \midrule
    CBAD                          & 0.970                  & 0.390            & 0.674                & 0.547                \\ \midrule
    CSD                           & 0.982                  & 0.788            & 0.941                & 0.871                \\ \midrule
    FCBAD                         & 0.994                  & 0.841            & 0.969                & 0.958                \\ \midrule
    GM-RX                          & 0.957                  & 0.546            & 0.557                & 0.219                \\ \midrule
    MD-RX                          & 0.986                  & 0.817            & 0.945                & 0.900                \\ \midrule
    RX                            & 0.985                  & 0.807            & 0.940                & 0.889                \\ \midrule
    SSRX                          & 0.747                  & 0.848            & 0.768                & 0.836                \\ \midrule
    WIN-RX                         & 0.749                  & 0.798            & 0.656                & 0.753                \\ \midrule
    AED                           & 0.979                  & 0.956            & 0.985                & \textbf{0.991}                \\ \midrule
    KIFD                          & 0.830                  & 0.995            & 0.992                & 0.989                \\ \midrule
    LSUNRSORAD                          & 0.994                  & \textbf{1.000}            & 0.980                & 0.910                \\ \midrule
    baseline HUE-AD & 0.962                  & \textbf{1.000}            & \textbf{0.993}                & 0.982                \\ \midrule
    mGE-AD & 0.927                  & 0.997            & 0.916                & 0.962                \\ \midrule
    UGE-AD   & \textbf{0.995}                  & 0.999            & 0.976                & 0.964 \\ \bottomrule               
    \end{tabularx}
    
\end{table}

\begin{table}[!htbp]
\centering
\caption{\label{tab:2}Comparison of ROC-AUC scores between various methods using the ABU airport dataset}

    \begin{tabularx}{\linewidth}{@{} Y YYYY @{}} \toprule
    
    \textbf{Methods}              & \textbf{ABU-airport-1} & \textbf{ABU-airport-2} & \textbf{ABU-airport-3} & \textbf{ABU-airport-4} \\ \midrule
    CBAD                          & 0.858                  & 0.633                  & 0.633                  & 0.886                  \\ \midrule
    CSD                           & 0.824                  & 0.826                  & 0.922                  & 0.934                  \\ \midrule
    FCBAD                         & 0.877                  & 0.880                  & 0.897                  & 0.986                  \\ \midrule
    GM-RX                          & 0.674                  & 0.590                  & 0.680                  & 0.453                  \\ \midrule
    MD-RX                          & 0.813                  & 0.841                  & 0.925                  & 0.896                  \\ \midrule
    RX                            & 0.822                  & 0.840                  & 0.929                  & 0.953                  \\ \midrule
    SSRX                          & 0.639                  & 0.857                  & 0.892                  & 0.644                  \\ \midrule
    WIN-RX                         & 0.774                  & 0.797                  & 0.949                  & 0.708                  \\ \midrule
    AED                           & 0.969                  & \textbf{0.994}                  & 0.957                  & \textbf{0.995}                  \\ \midrule
    KIFD                          & 0.940                  & 0.979                  & 0.961                  & 0.975                  \\ \midrule
    LSUNRSORAD                          & \textbf{0.972}                  & 0.959                  & 0.966                  & 0.959                  \\ \midrule
    baseline HUE-AD  & 0.970                  & 0.989                  & \textbf{0.973}                  & 0.988                  \\ \midrule
    mGE-AD & 0.859                  & 0.916                  & 0.920                  & 0.931                  \\ \midrule
    UGE-AD   & 0.933                  & 0.971                  & 0.948                  & 0.986 \\ \bottomrule
    
    \end{tabularx}
    
\end{table}

\begin{table}[!htbp]
\centering
\caption{\label{tab:3}Comparison of ROC-AUC scores between various methods using the ABU beach dataset}

    \begin{tabularx}{\linewidth}{@{} Y YYYY @{}} \toprule
    
    \textbf{Methods}              & \textbf{ABU-beach-1} & \textbf{ABU-beach-2} & \textbf{ABU-beach-3} & \textbf{ABU-beach-4} \\ \midrule
    CBAD                          & 0.571                & 0.603                & 0.664                & 0.885                \\ \midrule
    CSD                           & 0.982                & 0.911                & \textbf{1.000}                & 0.956                \\ \midrule
    FCBAD                         & 0.987                & 0.955                & \textbf{1.000}                & 0.948                \\ \midrule
    GM-RX                          & 0.986                & 0.701                & \textbf{1.000}                & 0.916                \\ \midrule
    MD-RX                          & 0.980                & 0.910                & \textbf{1.000}                & 0.955                \\ \midrule
    RX                            & 0.981                & 0.911                & \textbf{1.000}                & 0.954                \\ \midrule
    SSRX                          & 0.980                & 0.906                & 0.982                & 0.831                \\ \midrule
    WIN-RX                         & 0.957                & 0.962                & \textbf{1.000}                & 0.815                \\ \midrule
    AED                           & \textbf{0.997}                & 0.955                & \textbf{1.000}                & 0.979                \\ \midrule
    KIFD                          & 0.990                & 0.989                & 0.998                & 0.813                \\ \midrule
    LSUNRSORAD                          & 0.994                & 0.986                & \textbf{1.000}                & 0.975                \\ \midrule
    baseline HUE-AD  & 0.992                & \textbf{0.992}                & \textbf{1.000}                & 0.958                \\ \midrule
    mGE-AD & 0.962                & 0.980                & 0.981                & 0.935                \\ \midrule
    UGE-AD & 0.995                & 0.946                & \textbf{1.000}                & \textbf{0.980} \\ \bottomrule
    
    \end{tabularx}
    
\end{table}

\begin{table}[!htbp]
\centering
\caption{\label{tab:4}Comparison of ROC-AUC scores between various methods using the ABU urban dataset}

\begin{tabularx}{\linewidth}{@{} Y YYYYY @{}} \toprule

\textbf{Methods}              & \textbf{ABU-urban-1} & \textbf{ABU-urban-2} & \textbf{ABU-urban-3} & \textbf{ABU-urban-4} & \textbf{ABU-urban-5} \\ \midrule
CBAD                          & 0.391                & 0.155                & 0.613                & 0.598                & 0.573                \\ \midrule
CSD                           & 0.991                & 0.994                & 0.950                & \textbf{0.989}                & 0.969                \\ \midrule
FCBAD                         & 0.992                & 0.851                & 0.941                & 0.847                & 0.891                \\ \midrule
GM-RX                          & 0.760                & 0.549                & 0.778                & 0.765                & 0.402                \\ \midrule
MD-RX                          & 0.991                & 0.994                & 0.939                & \textbf{0.989}                & 0.970                \\ \midrule
RX                            & 0.991                & \textbf{0.995}                & 0.951                & \textbf{0.989}                & 0.969                \\ \midrule
SSRX                          & 0.928                & 0.989                & 0.748                & 0.786                & 0.791                \\ \midrule
WIN-RX                         & 0.927                & 0.909                & 0.783                & 0.787                & 0.881                \\ \midrule
AED                           & \textbf{0.998}                & 0.796                & \textbf{0.998}                & 0.840                & 0.984                \\ \midrule
KIFD                          & 0.938                & 0.917                & 0.992                & 0.978                & 0.984                \\ \midrule
LSUNRSORAD                          & 0.996                & 0.972                & 0.988                & 0.928                & 0.983                \\ \midrule
baseline HUE-AD & 0.993                & 0.972                & 0.996                & 0.985                & \textbf{0.990}                \\ \midrule
mGE-AD & 0.980                & 0.935                & 0.985                & 0.968                & 0.981                \\ \midrule
UGE-AD & 0.977                & 0.976                & 0.962                & 0.973                & 0.952 \\ \bottomrule    

\end{tabularx}
\end{table}

\begin{table}[!htbp]
\centering
\caption{\label{tab:5}Comparison of ROC-AUC scores between various methods using the Arizona dataset}

\begin{tabularx}{\textwidth}{@{} Y YYYYY @{}} \toprule

\textbf{Methods}              & \textbf{Arizona 1} & \textbf{Arizona 2} & \textbf{Arizona 3} & \textbf{Arizona 4} & \textbf{Arizona 5} \\ \midrule
CBAD                          & 0.510                       & 0.374                       & 0.556                       & 0.742                        & 0.648                        \\ \midrule
CSD                           & 0.335                       & 0.410                       & 0.601                       & 0.264                        & 0.645                        \\ \midrule
FCBAD                         & 0.367                       & 0.432                       & 0.662                       & 0.350                        & 0.727                        \\ \midrule
GM-RX                          & 0.576                       & 0.271                       & 0.457                       & \textbf{0.760}                        & 0.423                        \\ \midrule
MD-RX                          & 0.334                       & 0.410                       & 0.602                       & 0.271                        & 0.648                        \\ \midrule
RX                            & 0.334                       & 0.410                       & 0.602                       & 0.262                        & 0.647                        \\ \midrule
SSRX                          & 0.564                       & 0.578                       & 0.661                       & 0.458                        & 0.682                        \\ \midrule
WIN-RX                         & 0.534                       & 0.507                       & 0.629                       & 0.437                        & 0.596                        \\ \midrule
AED                           & 0.329                       & 0.326                       & 0.492                       & 0.307                        & 0.504                        \\ \midrule
KIFD                          & 0.580                       & 0.567                       & 0.671                       & 0.663                        & 0.780                        \\ \midrule
LSUNRSORAD                          & 0.208                       & 0.369                       & 0.501                       & 0.187                        & 0.499                        \\ \midrule
baseline HUE-AD & 0.248                       & 0.465                       & 0.551                       & 0.216                        & 0.573                        \\ \midrule
mGE-AD & 0.427                       & 0.290                       & 0.622                       & 0.382                        & 0.639                        \\ \midrule
UGE-AD & \textbf{0.671}                       & \textbf{0.695}                       & \textbf{0.925}                       & 0.458                        & \textbf{0.905} \\ \bottomrule

\end{tabularx}
\end{table}


\acknowledgments 
 
This research was sponsored by the Army Research Laboratory and was accomplished under Cooperative Agreement Number W911NF-21-2-0294 with Eddie L. Jacobs as PI. The views and conclusions contained in this document are those of the authors and should not be interpreted as representing the official policies, either expressed or implied, of the Army Research Office or the US Government. The US Government is authorized to reproduce and distribute reprints for Government purposes notwithstanding any copyright notation herein. 

This material is based upon work supported by the National Science Foundation while Lan Wang was serving at the National Science Foundation. Any opinion, findings, and conclusions or recommendations expressed in this material are those of the author(s) and do not necessarily reflect the views of the National Science Foundation.

\bibliography{report} 
\bibliographystyle{spiebib} 

\end{document}